# A Search for Peculiar Objects with the NASA Orbital Debris Observatory 3-m Liquid Mirror Telescope

Rémi A. Cabanac
cabanac@phy.ulaval.ca
Département de Physique, Université Laval, Ste-Foy, QC,
G1K 7P4, Canada
tel. (418) 656-2131 ext 6192, fax (418) 656-2040

Ermanno F. Borra
borra@phy.ulaval.ca
Département de Physique, Université Laval, Ste-Foy, QC,
G1K 7P4, Canada
tel. (418) 656-7405, fax (418) 656-2040

Mario Beauchemin
mabeauch@ccrs.nrcan.gc.ca
Natural Resources Canada, 588 Booth Str., # 411, Ottawa, ON
K1A 0Y7, Canada
tel. (613) 943-8829, fax. (947)-1406

**Abstract**.

The NASA Orbital Debris Observatory (NODO) astronomical survey uses a transit 3-m liquid mirror telescope to observe a strip of sky in 20 narrow-band filters. In this article, we analyze a subset of data from the 1996 observing season. The catalog consists of 18,000 objects with 10<V<19 observed in 10 narrow-band filters ranging from 500 nm to 950 nm. We first demonstrate the reliability of the data by fitting the Bahcall-Soneira model of the Galaxy to the NODO magnitude counts and color counts at various galactic latitudes. We then perform a hierarchical clustering analysis on the sample to extract 206 objects, out of a total of 18,000, showing peculiar spectral energy distributions. It is a measure of the reliability of the instrument that we extract so few peculiar objects. Although the data and results, per se, may not seem otherwise particularly remarkable, this work constitutes a milestone in optical astronomy since this is the first article that demonstrates astronomical research with a radically new type of mirror.

**Subject headings:**

instrumentation: miscellaneous, surveys, stars: peculiar, Galaxy: general, galaxies: peculiar, quasars: general.

## 1. INTRODUCTION

It has been known for centuries that the surface of a liquid spinning in a gravitational field has a parabolic shape. In principle, a spinning container filled with a reflecting liquid such as mercury could then be used as the primary mirror of a telescope. In practice, the concept never was taken seriously by astronomers for a variety of reasons, foremost of which are the facts that past experiments met with difficulties and, especially, that the mirror cannot be tilted and hence cannot track like a glass mirror. In recent times, the concept has been resurrected by Borra (1982) who argued that modern technology renders liquid mirrors useful to astronomy. Since that article, several developments have shown that the concept is viable. Optical tests showed that a 1.5-m diameter liquid mirror (LMT) was diffraction-limited (Borra et al. 1992): This article gives a wealth of information on the basic LM technology. This was followed by tests of a 2.5-m diameter liquid mirror (Girard & Borra (1997)) that also showed diffraction limited performance. Independent optical tests of a liquid mirror by Ninane & Jamar (1996) have confirmed diffraction-limited performance. Deep-sky imagery, obtained with the UBC-Laval 2.7-m LMT, has been demonstrated by Hickson et al. (1994).

Liquid mirrors are interesting in other areas of science besides Astronomy. For example, the University of Western Ontario has built a Lidar facility that houses a 2.65-m diameter liquid mirror as receiver (Sica et al. 1995). This telescope has now been operated trouble free for 5 years in the harsh Canadian climate, demonstrating the robustness of LMTs. The telescope is however not an imager. A Lidar facility has also been built and operated in Alaska by the University of California at Los Angeles. The most successful imager is the NODO 3-m diameter LMT (http://www.sunspot.noao.edu/Nodo/nodo.html) built by NASA (Potter & Mulrooney 1997, Mulrooney 1998) to monitor space debris. The observatory has operated for a few years. NODO has also been used to obtain a large number of nights of astronomical data (Hickson & Mulrooney 1997). Liquid mirrors have also interesting engineering applications (Thibault, Szapiel & Borra, 1998).

Although now somewhat outdated in this rapidly evolving field, a review paper (Borra 1995) elaborates on many aspects of Liquid Mirrors and Liquid Mirror Telescopes (including the all-important issue of the field of regard). Useful reference papers on LM technology and related issues have been written by Hickson and collaborators, ( e.g. Hickson et al. (1993), Gibson & Hickson (1992a, 1992b)).

While optical shop tests are needed to evaluate the intrinsic quality of the mirror and pretty pictures are reassuring, members of the astronomical community have repeatedly emphasized through personal communications that "the proof of the pudding is in the eating"; hence at some stage it is necessary to demonstrate scientific research. A previous article (Content et al. 1989) published astronomical research with a LMT but it used a small make-shift telescope with rudimentary instrumentation. The present paper demonstrates astronomical research with data obtained with a professional-quality 3-m LMT and is another important milestone in the demonstration of LMTs as Astronomical instruments.

**2. DATA DESCRIPTION**

The primary data set consists of 34 nights of narrow-band observations taken during the spring of 1996. Each night uses a single narrow-band filter selected among 11 bands (figure 1). The observations were taken with the NASA Orbital Debris Observatory (NODO) and are a subset of those described by Hickson & Mulrooney (1997). However, although we used the same primary images, we did our own independent data reduction with different algorithms and software, as described in the next section. The NASA Orbital Debris Observatory uses a 3-m diameter transit LMT. It is equipped with a 2048x2048 thick CCD camera continuously read-out at sidereal speed in the Time Delayed Integration mode (TDI). Tracking is thus done electronically by stepping, on the CCD, the pixels in the East-West direction at the sidereal rate. The integration time per night per object (97 seconds) is given by the time it takes an object to cross the 20-arcmin field of view. Longer integration times could be achieved by coadding data from different nights; however, in this work, we did not do any coadding. A typical night of observation gives a continuous band of observation which is divided, for convenience, in 280 20X20 arcminutes frames at constant declination, and increasing right ascension for a total of about 30 square degrees. The declination is given by the latitude of the observatory (32.96 deg). Each frame takes 8 Mb for a total of 2.2 Gb per night of observation.

The primary data has been described by Hickson & Mulrooney (1997) and evaluated by them so that there is no need to repeat their work here. Figure 2 gives an overview of the 27 nights of data that we examined and illustrates the robustness of the system. Every night of observation is labeled with the central wavelength of the bandpass used. The abscissa gives the time of observation converted to right ascension (top) as well as galactic latitude and galactic longitude (bottom). The timeline of a given night of observation is colored white whenever the images were deemed to be usable. The line is colored gray whenever the data was found to be unusable for whatever reason, including clouds. At the right of the timeline a brief comment is made. We can readily see that the majority of the timelines are white. The gray-shaded portions of the timelines are usually due to either clouds, to twilight or to cooling problems with the CCD detector which was too warm, hence had excessive dark counts. Although most of the nights were reasonably clear, some of them were definitely not photometric. In order to limit possible errors in the photometry from the beginning to the end of a night, we chose to extract the information from the best 11 nights, one per filter. Figure 1 gives the transmission curve of the filters , the quantum efficiency of the CCD (Hickson & Mulrooney 1997), and the night sky emission at Kitt Peak and Observatoire de Haute-Provence (Massey et al. 1990).

## 3. DATA REDUCTION

The entire data set was roughly 60 Terabytes in size. In order to extract useful information from this huge amount of data we wrote a set of C routines optimized for the format of the images. Each night was cut into about 280-300 2kx2k frames with an overlap of one arcmin between them. The set of automated C routines reduced the data according to the following steps:

**3.1 Image analysis**

0. Preprocessing was already performed on our images by Hickson & Mulrooney (1997) who removed the dark counts and subtracted the bias. We did not flat field our frames since driftscanning gives a very flat response. In the East-West direction, variations are near zero and we would not expect variations of more then 1% in addition to the gradient in the North-South direction.

1. Remove bad pixels and bad columns and average the gap using the nearest pixels and columns. This step is quite straightforward once one knows the coordinates of the faulty pixels. The CCD matrix was good except for one two-pixel column.
2. Convolve the image with a 3x3 pix gaussian kernel (FWHM = 2.35 pix = 1.4 arcsec). This reduces sky noise when present and increases the number of bits in the sky histogram. Sky flux was not always available because the CCD was not sensitive enough in the blue (fig. 1) and/or the integration time was too small. As a result, the noise was not sky-dominated for the blue filters ($\lambda < 6000$Å) but rather electronics-dominated. This did not have dramatic effects in the detection process, but it prevented us from measuring the actual magnitude limit of the sky and total efficiency of the system. Sky values and standard deviations are then measured and computed in a mosaic of 32x32 elements, each of them 64x64-pix large. The size of the mosaic elements is optimized to get the finest possible sky value within a region where the large-scale sky variation is lower then 1 $\sigma$.
3. Detect all pixels above 2 $\sigma$ of sky value (from the 32x32 mosaic) and identify groups bigger then 9 contiguous pixels and tag them as objects. This 2-$\sigma$ threshold, is our primary criterion to distinguish real objects from false detections. The 9 pixels can form any shape as long as they are contiguous. Hence, the total signal-to-noise ratio of the faintest detection is 6. This threshold is actually low since it allows 50% to 80% of all detections to be false. False detections are later eliminated by merging the nights.
4. Measure centroid, axial ratio, angle, Kron radius and growth curve from 2 to 10 pixels around every objects. The centroid is computed using the center of gravity of the complete distribution of all contiguous pixels. The axial ratio is calculated from all contiguous pixels without a priori subtraction. The position angle is the angle between the semi-major axis and the north-south direction. The axial ratio and the position angle are computed using the second moment of the spatial distribution (Hickson & Mulrooney 1997). The Kron radius is the isophotal radius defined in Kron (1980). The growth curve was computed over a circular disk between r = 2 pix and r = 10 pix by steps of 1 pix. Most of the objects are stars, thus point-like, and the axial ratio and position angles are usually sensitive to the wings of the PSFs near the edge of the field where the TDI sidereal velocity is either too small (north) or too large (south). Where Hickson & Mulrooney (1997) use a corrected isophotal flux, we use a simple aperture photometry with a 3 arcsec diameter circular aperture.

**3.2 Photometric and astrometric calibration**

The photometric characteristics of the data, albeit reduced with different software and algorithms, have been described by Hickson & Mulrooney (1997) and evaluated by them; therefore we will not repeat it here. We shall, however, describe briefly our own data reductions.

To make a zero-point calibration of the nights we searched astronomical databases to find calibrated standard stars in our strip of sky. We only found two blue stars, one at each end of the strip, with featureless spectra calibrated in AB magnitudes (Oke 1990). In practice, only the star at the beginning of the night could be used since the one at the end of the night was too bright and saturated the CCD in some filters. Tests show that the zero-point calibration does not vary by more the 0.3 mag from the beginning to the end of the nights that we chose. Fortunately, the search for peculiar objects, which is the primary scientific objective of this article does not need an accurate zero-point calibration.

After the primary reduction and photometric measurements were done, the next important step was to carry out an accurate astrometry over the whole night. The coordinates of the objects are the principal criterion used for identification and merging of objects from one night to another. The astrometry is calibrated in two steps, first we use analytic transformations from instrumental coordinates to J2000 equatorial coordinates. This allows to correct for precession, nutation, and aberration. Because LMTs are zenith instruments, they are particularly sensitive to astronomical aberration (up to 20 arcsec from the beginning to the middle of the night). For the second step we wrote a fine-tuning program, using standards from the United States Naval Observatory astrometric survey (Monet 1997). We corrected the coordinates frame by frame comparing all matched stars (usually over 50/frame, up to 500). The final astrometry is good within 0.5 arcsec. We merge the nights into one catalog containing the coordinates and the magnitudes of all objects. We only included in the catalog the objects detected in every filter. Figure 3 shows the magnitude counts in all of the filters of the final catalog. It is difficult to estimate the completeness of the catalog because the quantum efficiency of the detector varies considerably over the 4,000 Å spectrum. In particular, the CCD is not very sensitive in the blue; hence the catalog is biased toward faint blue objects.

## 4. REPRODUCING THE KNOWN: COMPARISONS TO IMAGES FROM THE PALOMAR OBSERVATORY SKY SURVEY AND TO MODELS OF THE GALAXY

Before a new instrument is used to carry out original research it must demonstrate that it can reproduce known results. We therefore first make a qualitative comparison to images from the Digitized Sky Survey II. We then compare our star counts with standard models of the Galaxy. Hickson & Mulrooney (1997) also give an evaluation of the data which complements ours.

**4.1 Comparing the LMT images to images from the Palomar Observatory Sky Survey.**

Eye inspections of a very large number of frames in all filters and at all times of the nights indicate that the images are well behaved. We also visually compared in greater details selected frames to the corresponding images of the DSS II. For example, figure 4 shows such a comparison between an unfiltered (equivalent to a broad-band red filter) deep image of the NODO survey (top) and the DSS II image (bottom). One may appreciate the similarity between the CCD image and the digitized photographic plate. NODO narrow-band images are generally deeper in red than DSS II, and shallower in blue. Hickson & Mulrooney (1997) gives limiting magnitudes and associated magnitude errors for all filters.

**4.2 Comparing star counts and colors to models of the galaxy**

We next compare star counts, extracted from our database to the well-established Bahcall-Soneira models of the Galaxy. Bahcall (1986) proposed a simple analytical model of the Galaxy containing a spheroid and a disk component and wrote a code computing the magnitude counts and number counts by simply summing the contributing stellar populations in the line-of-sight of the observer. The Bahcall-Soneira models reproduce past observations remarkably well (Bahcall et al. 1985, Lasker et al. 1987), and they can be considered as reliable standards for the distribution of the stars in the Galaxy in a range of magnitude B < 20. Recent studies by Santiago et al. (1996) using HST data, Bath et al. (1996)

using the APS catalog, and Malkov & Smirnov (1994) using the Guide Star Catalog show similar agreements in star counts, but some discrepancies in color counts attributed to a thick disk component. We used the Galaxy parameters given in Table 1 (Bahcall 1986).

The code was found at http://www.sns.ias.edu/jnb/galaxy/html, courtesy of J. Bahcall. We did not change the normalization of its computed results. We use directly the output of the program to produce count and color curves at three different galactic latitudes for different magnitude intervals. We transform the standard B and V magnitudes to AB magnitudes in our narrow-band filters at 500 and 700 nm. The transformation equations are calculated using the standard spectra of the blue star Hz 21 from Oke (1990) for the zero-point calibration, and the Gunn & Stryker catalog of stellar spectra (1983) for the color calibration. Least square fits yield the transformation equations:

$$M50 = V + 0.385 (B-V) - 0.033, \qquad (1)$$

and for $-0.4 < B-V < 0.9$:

$$(M50-M75) = 1.36 (B-V) - 0.40 \qquad (2)$$

for $B-V > -0.9$:

$$(M50-M75) = 3.00 (B-V) - 2.1 \qquad (3)$$

where M50 and M75 are the AB magnitudes in our catalog, and B and V are the standard bands used by Bahcall and Soneira. Figures 5 and 6 show the data and models for three different galactic latitudes and longitudes where extinction is less than 0.01 (Burstein & Heiles 1982). In the models, the counts are sensitive to the giant branch population of the spheroid component of the Galaxy. The metal poor cluster M15 gives the best fits towards the north galactic pole, and the rich cluster 47 Tucanae gives the best fits for lower galactic latitudes. This is what one expects if spheroid stars are more metal-poor than disk stars. Our data is thus consistent with past observations. The models fit well the counts for four different magnitude intervals, although photometric errors make the comparison less reliable for V>19. There is a small shift in the color peak near the plane of the galaxy (corresponding to the end of the nights). Our data are plotted as they come straight out of the database. There is no renormalization in neither counts nor colors. Hence, because we only use a single photometric standard at the beginning of the nights, the shift is probably due to extinction variations from the beginning to the end of the nights. It is a testimony to the photometric qualities of the nights that the comparisons between models and data are as good as shown in figures 5 & 6.

**5. A SEARCH FOR PECULIAR OBJECTS WITH A HIERARCHICAL CLUSTERING ANALYSIS**

**5.1 Simulations**

We created a database of artificial spectral energy distributions of different kinds of astronomical objects, as they would look like seen through our filters, by convolving higher resolution spectra, obtained from the literature, with the 10 filters shown in figure 1. For stars, we used the Gunn & Stryker catalog of stellar spectra . For the quasar we used a composite "typical" spectrum. Unfortunately quasar spectra display a broad variety of spectra and a "typical" spectrum is a poor approximation. Figure 7 shows our library spectra for E , Sc and Irr galaxies at the different redshifts as well as our "typical" QSO at different redshifts. The galactic spectra have not been corrected for evolution. We assume that whatever might happen in the stellar population of the giant galaxies, it will only change the continuum by small amounts for

z<0.5. This is probably a crude approximation for dwarf irregulars and galaxies that experience starbursts, but this assumption is not critical, since the purpose of the simulation is not to test evolutionary models but to get an idea of what galaxy spectra, seen through our filters, would look like at different redshifts. Figure 7 shows that Irregular galaxies and Sc galaxies as well as QSOs have striking features especially at high redshift. Since the filters do not cover the whole range between 500nm to 950nm, emission lines are not always detected. For instance, while the QSO shows strong lines at z=3.9, because both Ly α (1216Å) and CIV (1549Å) fall in two of our filters, the same lines are no longer visible at z=4.1. Similarly, the usually conspicuous break at 4000Å in giant ellipticals (Cabanac et al. 1995) is almost invisible at z=0.3, in figure 7. The dwarf irregular galaxies show an interesting continuum inversion at high redshift (z>0.45). This simulation gives another tool to extract information from the peculiar object spectra given by the HCA (next section).

**5.2 Hierarchical Clustering Analysis**

In the rest of the paper we will refer to flux distributions of the 10 filters as spectra, although we do not uniformly cover the entire range between 4000 Å and 9000 Å (see figure 1) and thus do not have the complete low-resolution spectra of the objects. Notwithstanding these limitations, our catalog carries more information than an equivalent BVRI photometry catalog. Here, we present a search for peculiar objects on a subset of 18,000 bright objects using a hierarchical clustering analysis (hereafter HCA). The method is based on the technique used by Beauchemin et al. (1993) (also Beauchemin et al. 1991) using HCA code from Murtagh & Heck (1987). A comprehensive description of the technique can be found in the previous cited papers and references therein. HCA is a multi-dimensional clustering technique that uses a minimum variance criterion to segregate objects in groups. To perform HCA, we sliced the catalog into sub-samples of 2,000 objects each. Table 2 gives the beginning and the ending right ascensions, and total area for each sub-sample. The total area of each sub-sample decreases from the north galactic pole to the galactic plane as a result of an increasing density of objects. We sliced our catalog for two reasons: First, HCA is time-consuming and the CPU cost increases as the square of the number of objects. Second, each sub-sample covers a shorter time length then the full night, slicing the data set reduces the impact of photometric variations. Our analysis indicates that the impact of varying atmospheric extinction is negligible within our time subsamples. We excluded the flux of the last filter (#95) from the objects, because the combined low sensitivity of the CCD and high sky brightness (figure 1) induced large photometric errors which falsely produced many peculiar objects. We did not select any stopping rule to choose the optimal number of clusters (Milligan et al. 1985) since our purpose was not to create a rigorous classification of objects but rather to group them according to a relative spectral morphology within a given number of clusters. The HCA output gives a hierarchical tree (or dendogram) with a maximum of 40 branches (or clusters) assembling similar objects. We may cut the dendogram at any of the 40 levels, corresponding to the number of clusters we want. To choose the number of clusters, we tested the dendograms with 5, 10, 20 and 40 clusters. We needed to isolate as many peculiar objects as possible from the rest of the subset. We concluded that the best number of clusters was 40. A smaller number of clusters did not leave enough free clusters for outliers. Figure 8 gives the result of HCA on the group 0. Median spectra of selected groups are plotted in figure 9.

For the HCA, the objects are normalized at 700 nm. We plot the median spectra with their true AB magnitudes.

All objects in clusters containing less than 10 objects were tagged as potentially interesting candidates. In this first iteration, we extracted 445 objects from the 18,000 objects of the catalog. We applied the HCA a second time on the sub-sample, using 10 clusters. This effectively classified the sub-sample by colors. Figure 10 shows the result of the HCA and median spectra of each cluster are shown figure 11. Figure 12 compares 5 observed stellar spectra to the simulations. The observed spectra reproduce well the simulated spectra, another testimony as to the reliability of our data.

In order to associate an approximate spectral type to each object, we fitted the 445 objects with the best fitted spectrum from a database of spectral energy distributions obtained from our spectral library (section 5.1). Even if our spectra were not rigorously corrected, this step permitted to isolate all spectra that looked stellar. We eliminated from our first sample of 445 objects all objects showing unambiguous M and K star features. Because cool stars are relatively rare at these magnitudes, the HCA classified a number of these normal stars as peculiar.

Because different wavelengths are observed on different nights, variable objects have unreliable spectra. Our spectra are also light curves! Note however that consecutive filters were not taken on consecutive nights so that time does not increase (nor decrease) from blue to red. Light variations will show up as spectral variations so that we should expect that a fraction of our peculiar objects are actually variable objects. This is almost certainly the case for stellar looking continua that show a deep absorption (see figure 13, #126, #38). Objects that have strong enough light variations show up as peculiar spectra and our catalog is probably heavily contaminated by these objects. We remove from our sample some of the most obvious cases.

Finally, we keep the remaining 206 objects and looked at their spectra and images for hints of peculiar features, emission peaks, halos, or companions. The list of these peculiar objects with their coordinates, AB magnitudes in filter 550 (essentially the V band) comments such as, suspected type, best-fitted stellar library spectra, striking spectral features is given in Table 3. A postscript figure containing all peculiar spectra is available at
http://wood.phy.ulaval.ca/lmt/abstracts/peculiar.spectra.ps.z (gzipped 230kb). Figure 13 illustrates what we mean by absorption lines (a) or (sa), emission lines (e) or (se), breaks (b) or (sb), variable (v), and inverted continuum (i) in the comments of table 3 ("s" in sa, se and sb stands for "strong"). Eyeball classification of spectral features is highly subjective since it depends on where the eye sets the continuum. Independent classification by two of us shows disagreements as to whether a feature is an absorption, an emission or a break. Consequently, the classification of a feature in Table 3 should be taken cautiously: the comments really only indicates that there is some spectral feature near that wavelength. Follow-up spectroscopy is needed. The following section is devoted to particularly interesting objects that deserve a spectroscopic follow-up.

### 5.3 Peculiar Objects

In this subsection, we comment on some of our most interesting objects.

**Optical counterparts of radio source : Objects #50, 122**

We checked whether some of our 206 peculiar objects coincide with known peculiar objects by cross-referencing them with the NASA/IPAC Extragalactic Database (NED). We find that two of them are optical counterparts of known radio sources. Figure 14 shows the red image of each object with superimposed contours of the Radio Source. The image comes from the Digital Sky Survey and the Radio contours come from the VLA FIRST Survey. Both optical images and radio contours are extracted from the SkyView image database. The radio sources, 87GB123819.7+332002 (obj#50; already known as an optical counterpart in NED) and 87GB162630.1 +325701 (obj#122; discovered by us) have steep radio spectral indices, implying non-thermal emission processes in the transparent radio lobes. Both optical objects show continuum radio emission at 1.4 Ghz. Extended emissions in the immediate vicinity of the objects are barely detectable on the DSS. The south lobe of the obj#50 seems to have an optical counterpart, and the obj#122 (galaxy) is surrounded by what looks like faint galaxies. Obj#122 might be of particular interest since the lobes have peculiar morphologies and dynamical studies could infer properties of the galaxy cluster.

**Gravitational lens candidates , close pairs and groups: Objects #48-49, 176-177**

We find two gravitational lens candidates(#48-49, 176-177), as well as several close pairs and groups are identified (#53-54, 40-119, 57-61-146, 153-157, 169-170, 129-134). The selection criteria are that the objects should have similar spectra and that their centroids should be separated by less than 10 arcsec for gravitational lens candidates, and 10 arcmin for close groups. Figure 15 plots the spectra of the two gravitational lens candidates. The close groups are probably stars with similar spectra, but they could also be remote members of galaxy clusters. At a redshift of 1, an angular distance of 10 arcmin is equivalent to a distance of 5 Mpc ($q_o=1/2$, $H_o=50 km.s^{-1}Mpc^{-1}$).

**Blue objects : Objects #1-47**

14 blue objects (#1-14) were detected with the colors of O or B stars, and 33 objects (#15-47, 185) with colors of A stars. They could be hot subdwarfs, white dwarfs or QSOs.

**Object #203: Galaxy cluster**

Finally, an interesting object was found while doing eyeball checking of a randomly selected sample of Table 3. Obj#203 is a red object (probably an elliptical galaxy) surrounded by a great number of faint extended objects. The DSS image reveals a large number of galaxies inside a radius of 10 arcmin around obj#203.

**Inverted spectrum objects**

These objects (e.g., figure 13, #150) are potentially quite interesting; however we find that the majority of them are very faint, so that the inversion may actually come from faulty sky subtractions. However, at least two (#14, #134) are bright enough that this explanation does not hold. We cannot exclude that they are variable objects; however the fact that increasing time does not correlate with increasing (nor decreasing) wavelengths, makes this somewhat unlikely.

## 6. CONCLUSION

The main scientific goal of this paper consists in a search for objects that have peculiar energy distributions. We extract spectra for 18,000 objects having $10<V<19$ in a 15-arcminute strip of sky extending from 12h30 to 18h00 for a total of ~18 square degrees. We then perform a hierarchical clustering analysis on the sample in order to extract objects showing peculiar spectral energy distributions. A blind HCA analysis finds 445 objects having peculiar energy distributions. Inspection of the energy distributions shows that some of these objects are actually cool stars so that they were removed from the list. Because different filters are used on different nights, variable objects show up as peculiar objects. We therefore removed from the list objects that are obvious variable stars as well as some objects that have false features caused by problems with the detector (e.g. cosmic rays). The final list contains 206 objects. Because we do not have redundant observations, there is a very high likelihood that most of these objects are actually variable objects, so that follow-up spectroscopy is needed to confirm their peculiarities. On the other hand, to the best of our knowledge, this survey is unique because of the combination of spectral resolution, relatively faint limits and area surveyed. It therefore has good potential to find rare objects.

Although the data and results, per se, may not seem particularly remarkable, this paper constitutes a milestone in optical astronomy since it is the first paper to demonstrate astronomical research with such a radically new type of optics. This is a first generation liquid mirror telescope and its instrumentation is actually not optimized for astronomical observations since the detector is a thick CCD with low quantum efficiency and the NODO 3-m telescope was designed to observe fast moving space debris so that the electronics generates a strong read-out noise. Still, the telescope demonstrated its robustness by consistently giving good images over a large number of nights. That the data is reliable is demonstrated by the fact that star counts and color histograms fit the Bahcall & Soneira model of the Galaxy. It is also demonstrated by the fact that the impartial HCA search only found 445 objects out of 18,000 in its first blind pass.

Considering the technology at the time of this writing, the low capital and operational costs of the first generation liquid mirror telescopes make them ideal specialized instruments for narrowly focused surveys where the limited field of regard is not a serious limitation. However, future advances in corrector designs could render the next generation LMTs far more versatile by greatly extending their fields of regard (Borra, 1995).

**Acknowledgments**

We thank Paul Hickson and Mark Mulrooney for providing an invaluable set of data. This work has been made possible by the use of the USNO-A astrometric Catalog (Monet 1997), the Digitized Sky Survey produced at the Space Telescope Institute under U.S. Gov. Grant NAG W-2166, the NASA IPAC/Extragalactic Database (NED) operated by JPL, Caltech, under contract with NASA, the SkyView image database developed and maintained under NASA ADP Grant NAS5-32068, at the High Energy Science Archive Research Center (HEASARC) at the GSFC. This work was supported by NSERC grants to EFB, and an NSERC fellowship to RAC.

**Figure captions**

**Figure 1**
It gives the typical quantum efficiency (QE) for a thick 2kx2k, the sky brightness from Kitt Peak, and the transmission curves (TC) of the filters.

**Figure 2**
Overview of a subset of the 1996 season of observation of the NASA Orbital Debris Observatory. The gray shades overlap non-optimal conditions of observation.

**Figure 3**
Plots the magnitude counts of the final catalog for objects detected in all colors. The completeness is poor beyond mag~18.5 in red and 17.5 in blue. This comes from the CCD which is not optimized for astronomical purposes. There are combined effects of saturation in red (mag<13) and poor efficiency in blue that strongly limit the linear range of detection.

**Figure 4**
Comparison between an image taken with the NASA Orbital Debris Observatory (top) and its POSS II counterpart (bottom). The field is 5 x 7 arcmin. RA = 12h08m, DEC = 33.0° (J2000.0).

**Figure 5**
Plots magnitude counts at three galactic latitudes: (l=86°, b=83°) (l=53°, b=47°) (l=56°, b=34°). The steps are the observed counts per square degree, and the dash-lines are the Bahcall-Soneira models.

**Figure 6**
Plots the observed color counts (in steps) at same latitudes than figure 5, for four intervals of magnitudes (V<14, 14<V<17, 17<V<18.5, 18.5<V<19.5) and their associated Bahcall-Soneira models of the Galaxy (in dash-lines). The different theoretical curves use different stellar populations for the spheroid component of the Galaxy.

**Figure 7**
Gives the result of HCA on group 0 (first 2000 objects) of the catalog. The histogram shows the number of objects in each cluster.

**Figure 8**
Plots of the median spectra for each cluster of the Fig. 7.

**Figure 9**

Shows the result of HCA for the 445 objects extracted from the 18,000 objects.

**Figure 10**
The plots show the median spectra for each of the 10 clusters of figure 8.

**Figure 11**
Show simulations of stellar spectra (B, F, G, K, M) filtered in our 10 bands, different types of galaxies (E, Sc, Irr) at various redshifts and a composite quasi stellar object spectra at various redshifts. It illustrates, the diversity of the possible objects (no evolution).

**Figure 12**
Gives 5 spectra and their best-fitted stellar spectra, from O8 to M8. All objects which showed obvious similarities with stars (e.g. M8) were removed from the peculiar object catalog.

**Figure 13**
Illustrates the comments given in Table 3. Absorption (a or sa), emission (e or se), break (b or sb), inverted continuum (i) and variable (v). the numbers are the same as in Table 3.

**Figure 14**
Images of two of the peculiar objects which are optical counterpart (line-of-sight) of two extragalactic radio sources. The gray scale images are DSSII images and the contours are FIRST 1.4Ghz flux.

**Figure 15**
Spectra of two gravitational lens candidates.

**Table titles**

**Table 1:**

Galaxy parameters for the model of the galaxy.

**Table 2:**

HCA sub-samples.

**Table 3:**

Peculiar objects

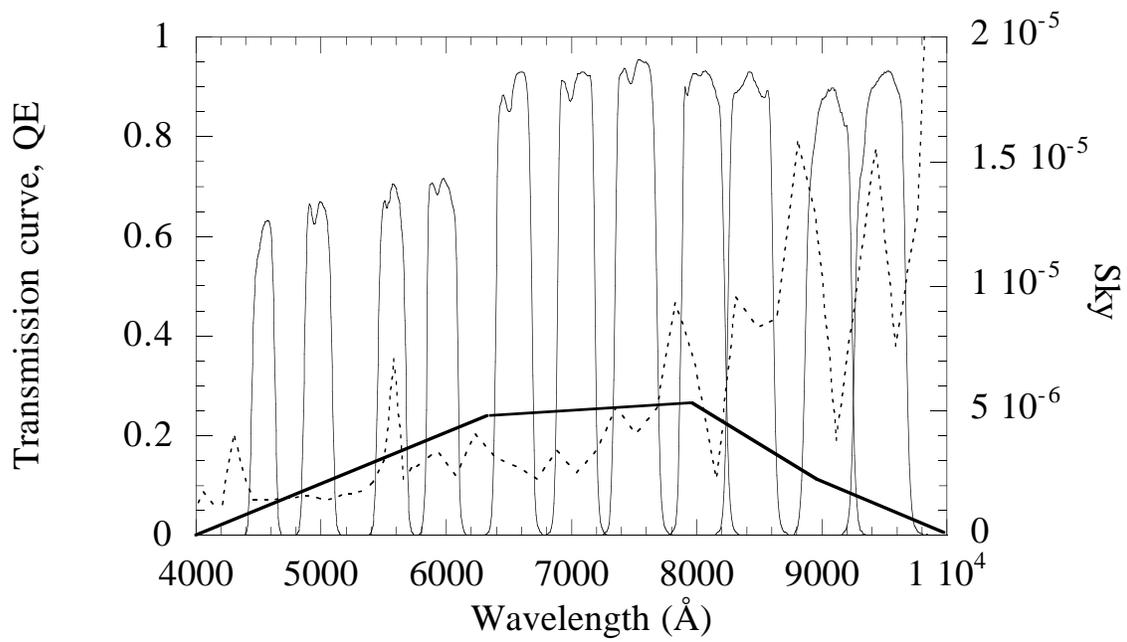

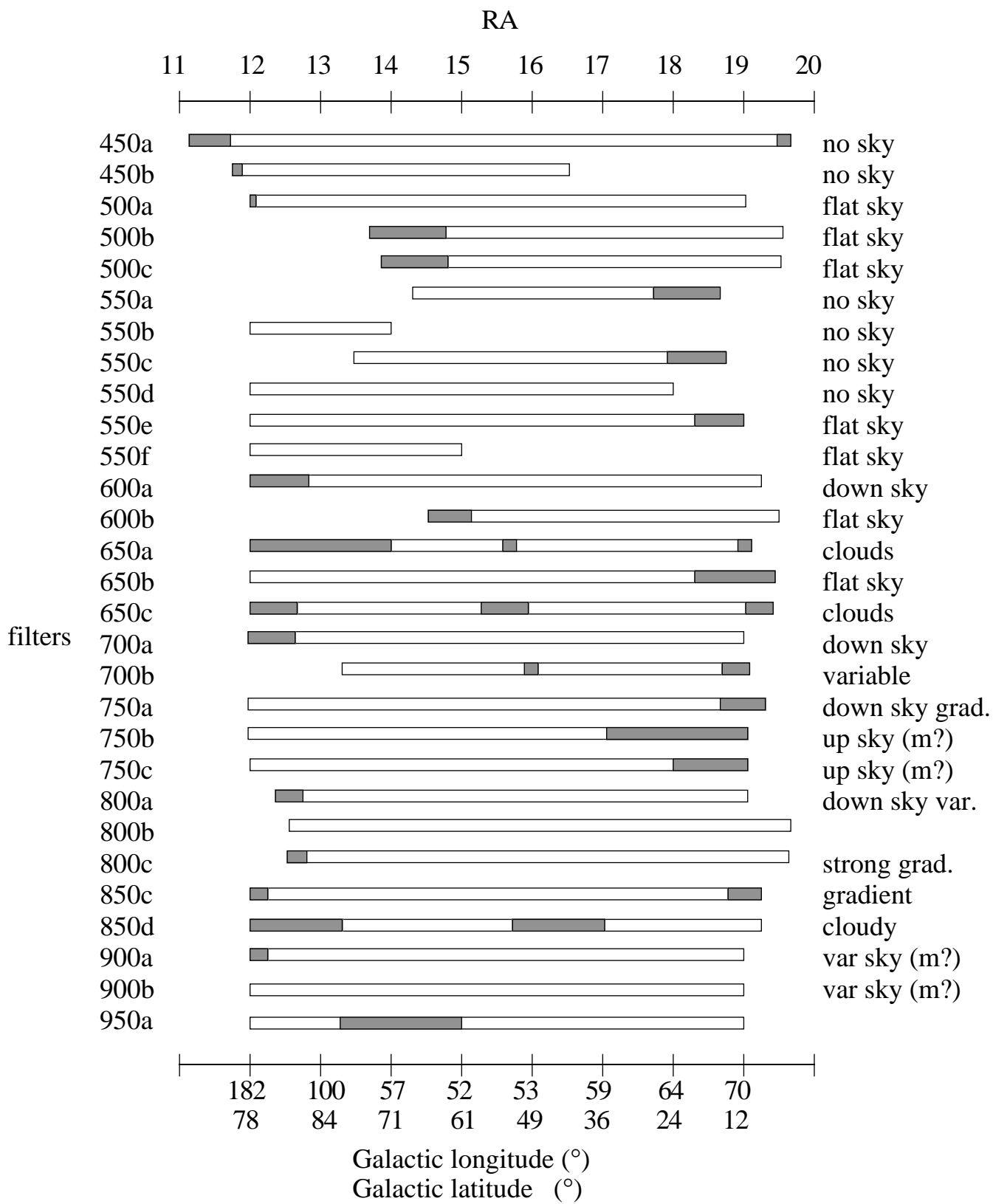

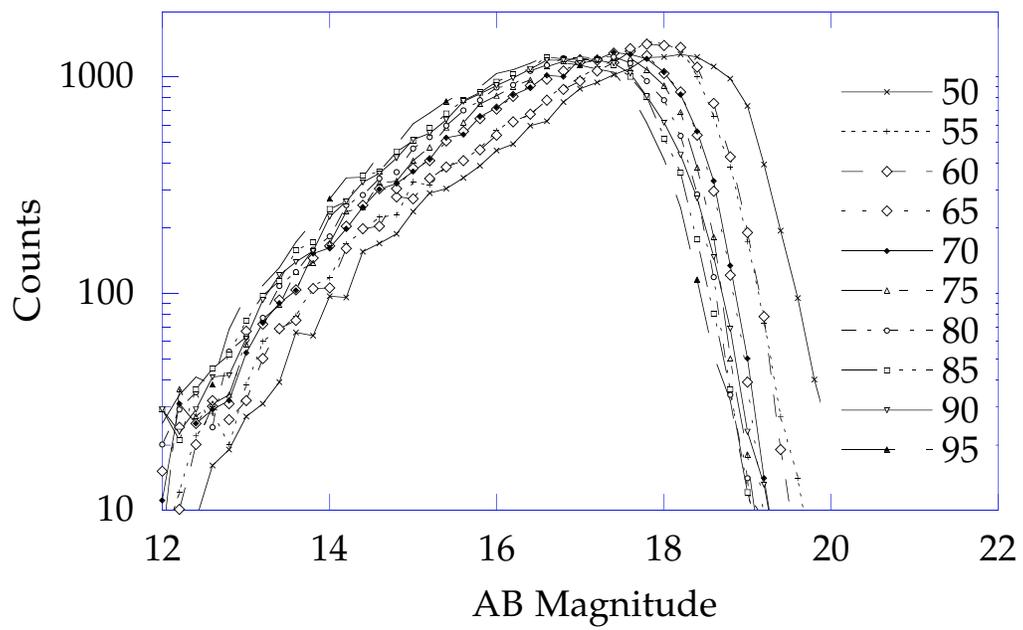

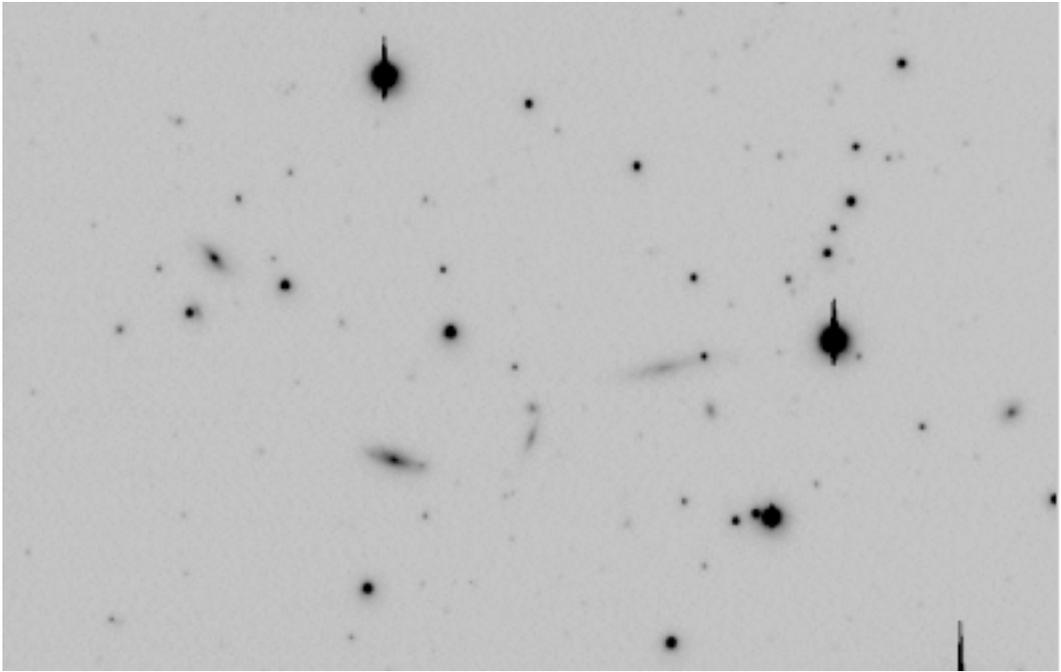
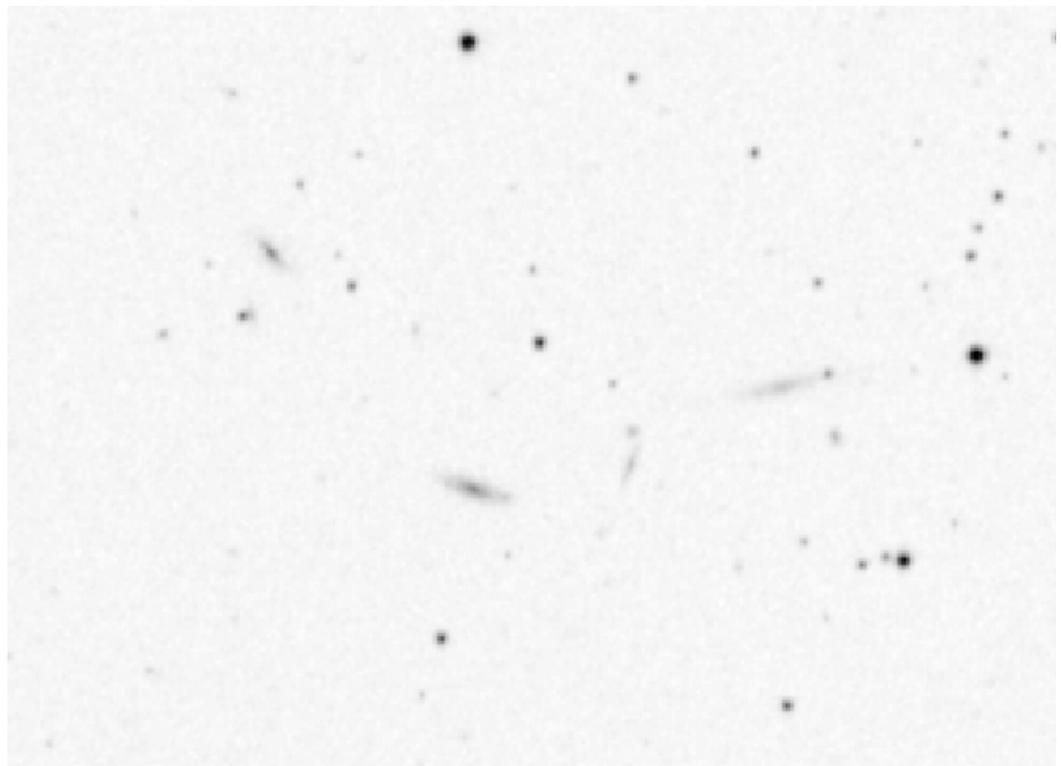

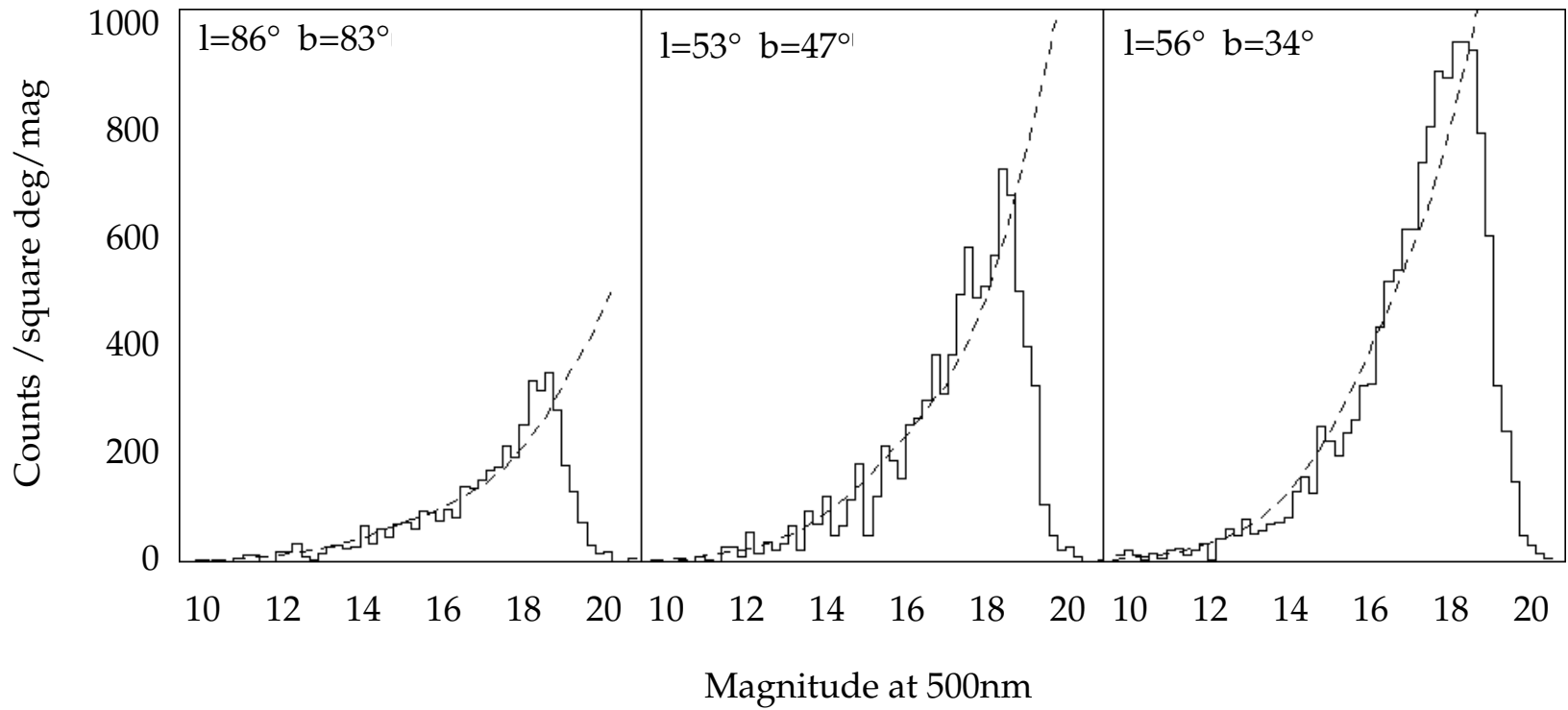

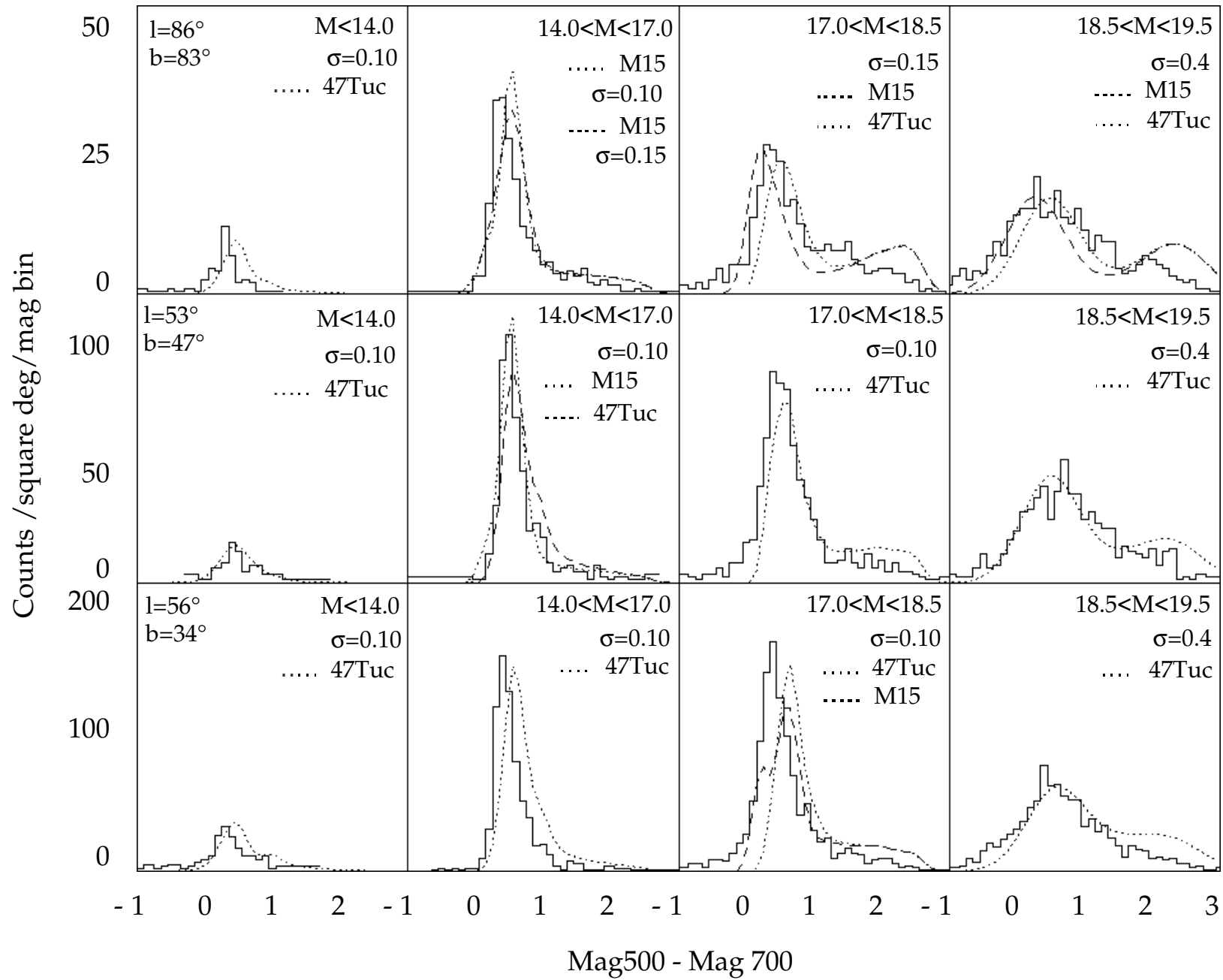

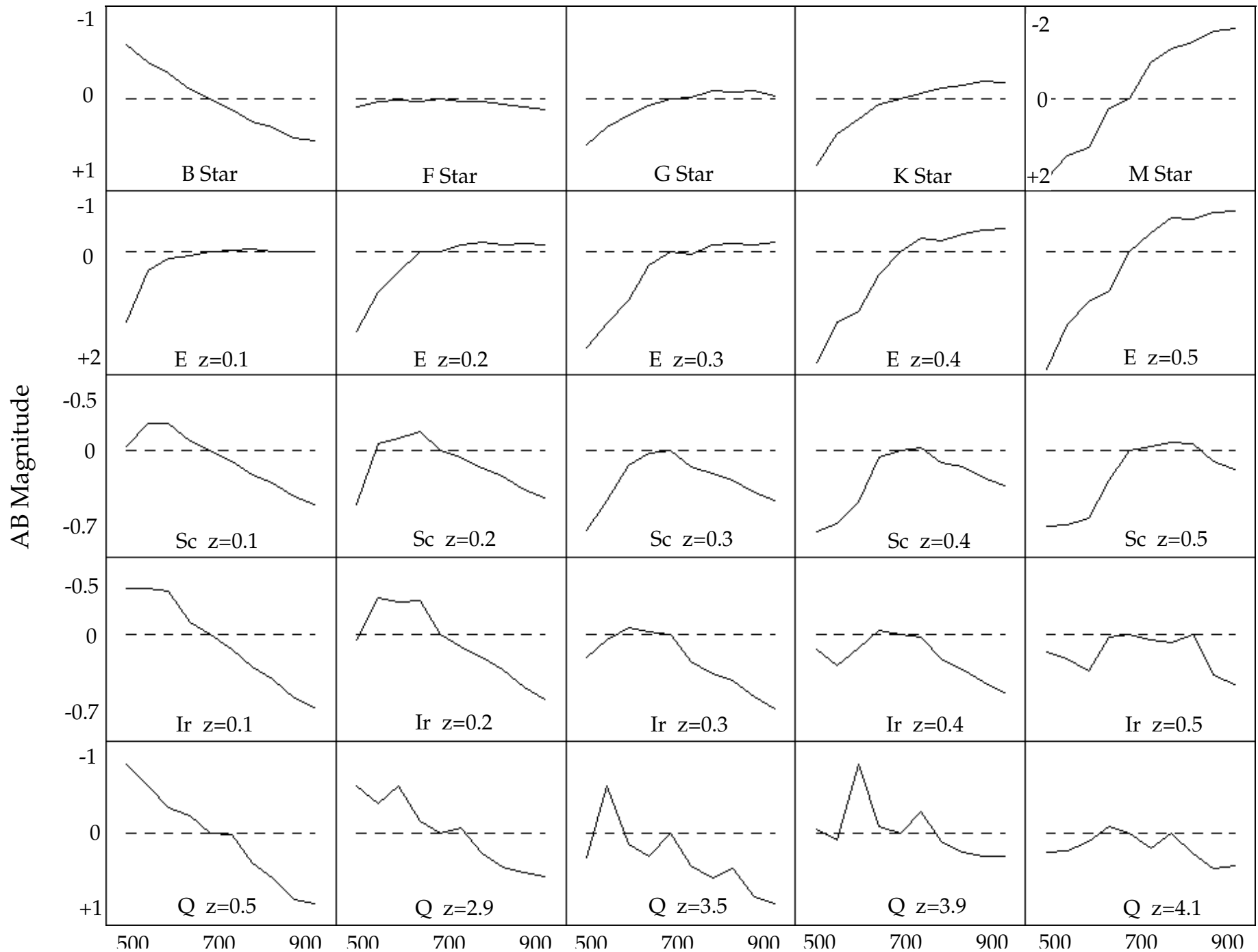

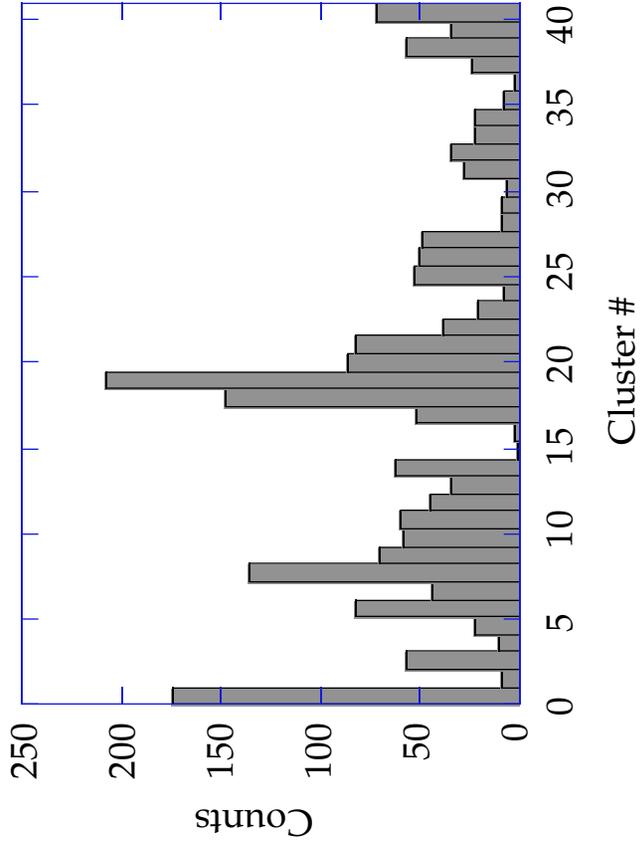

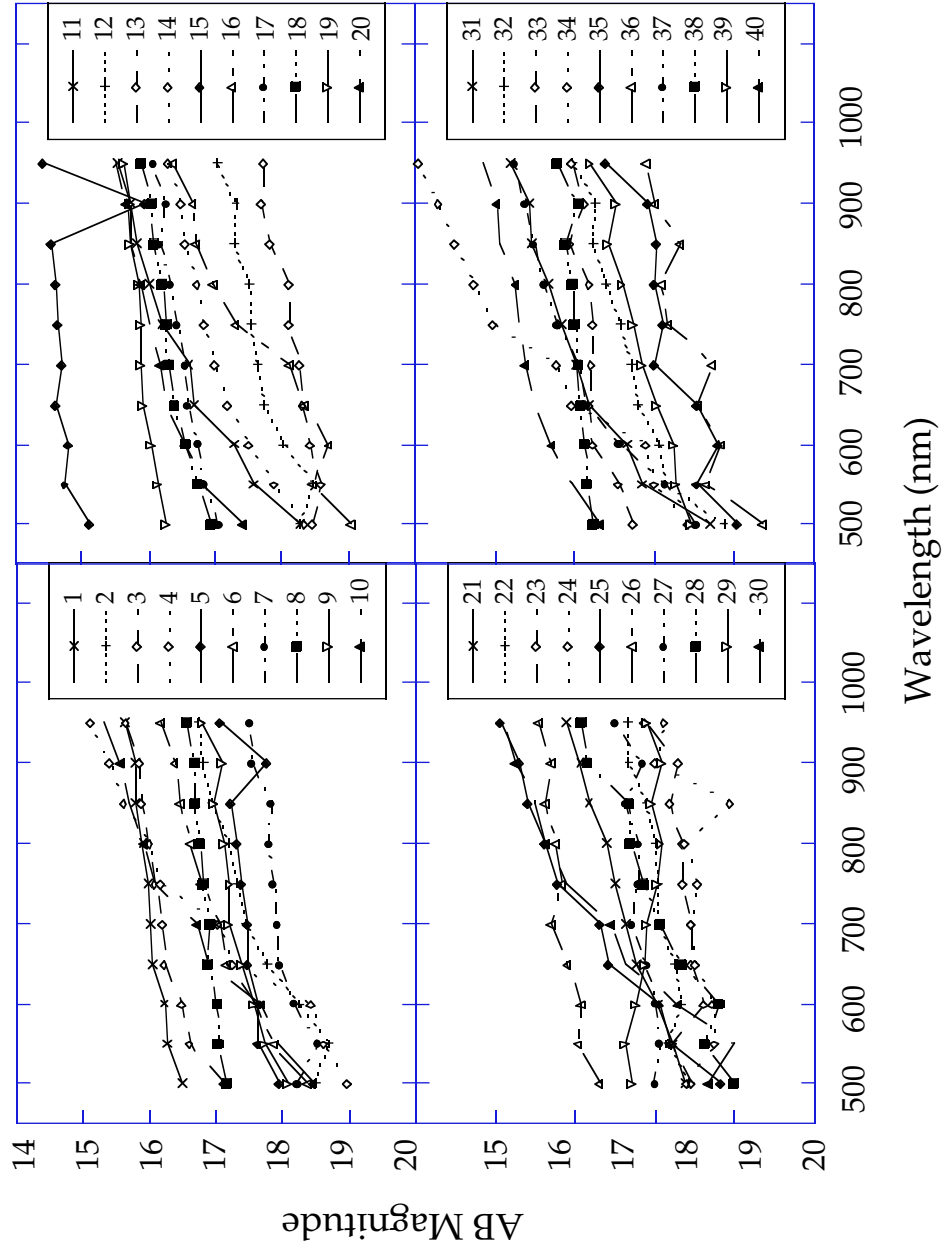

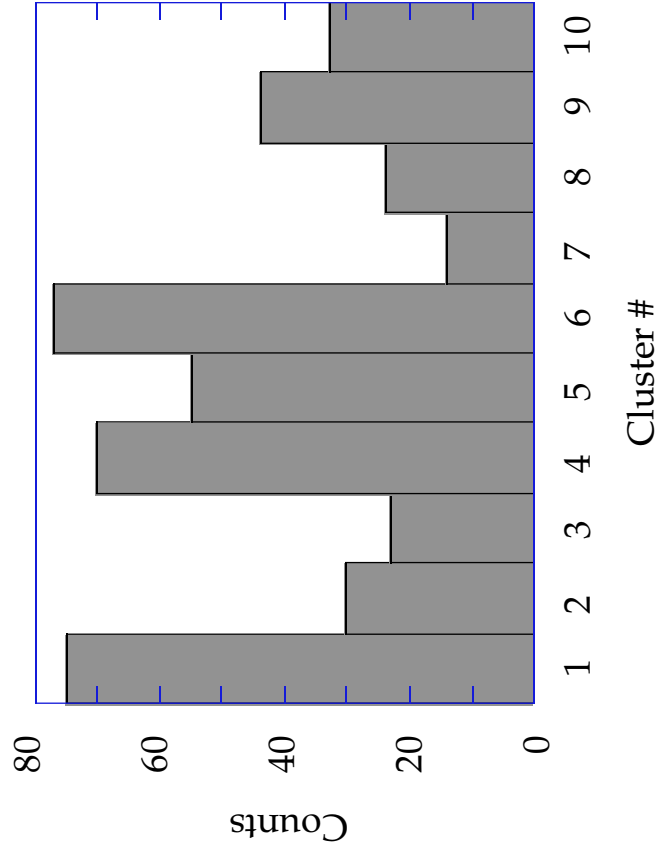

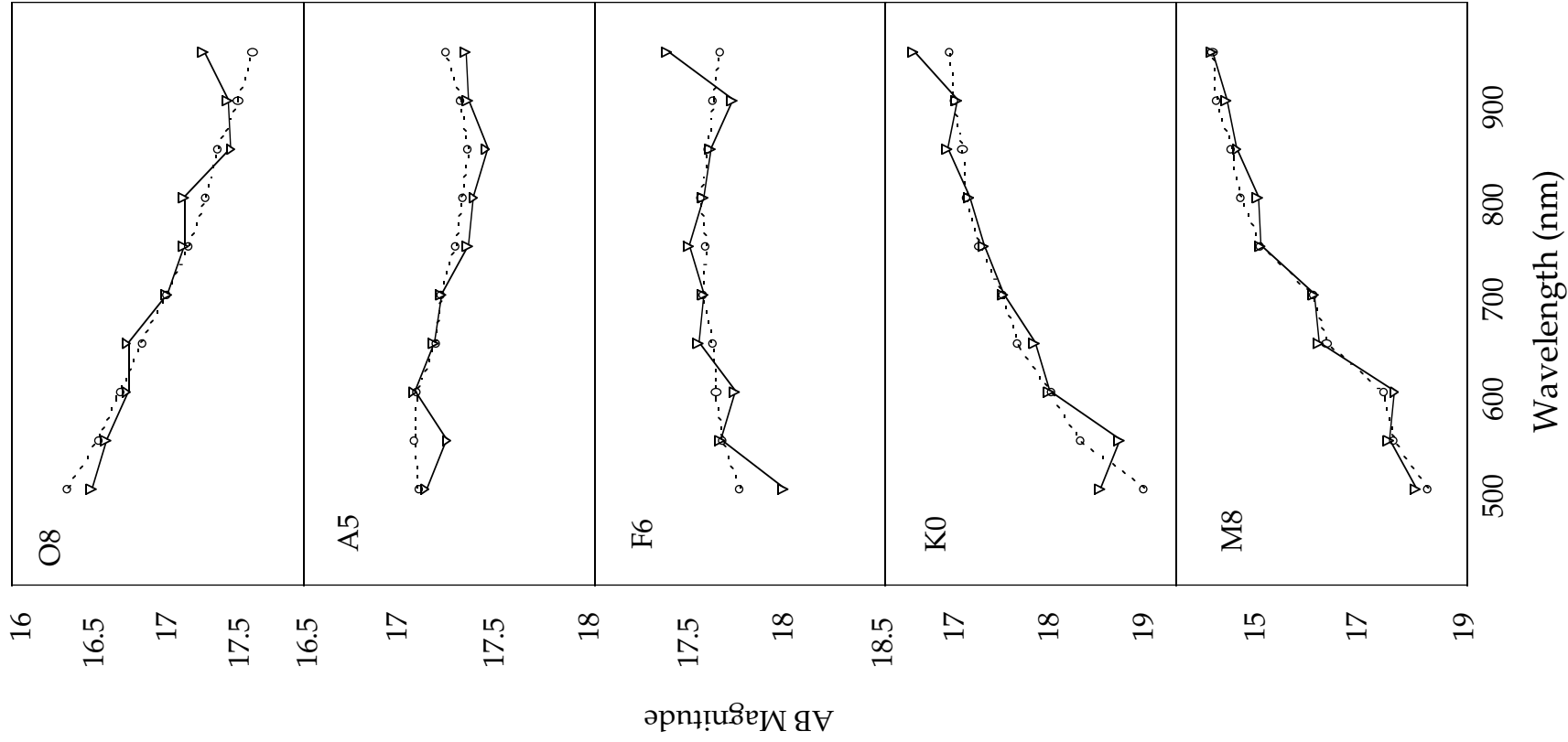

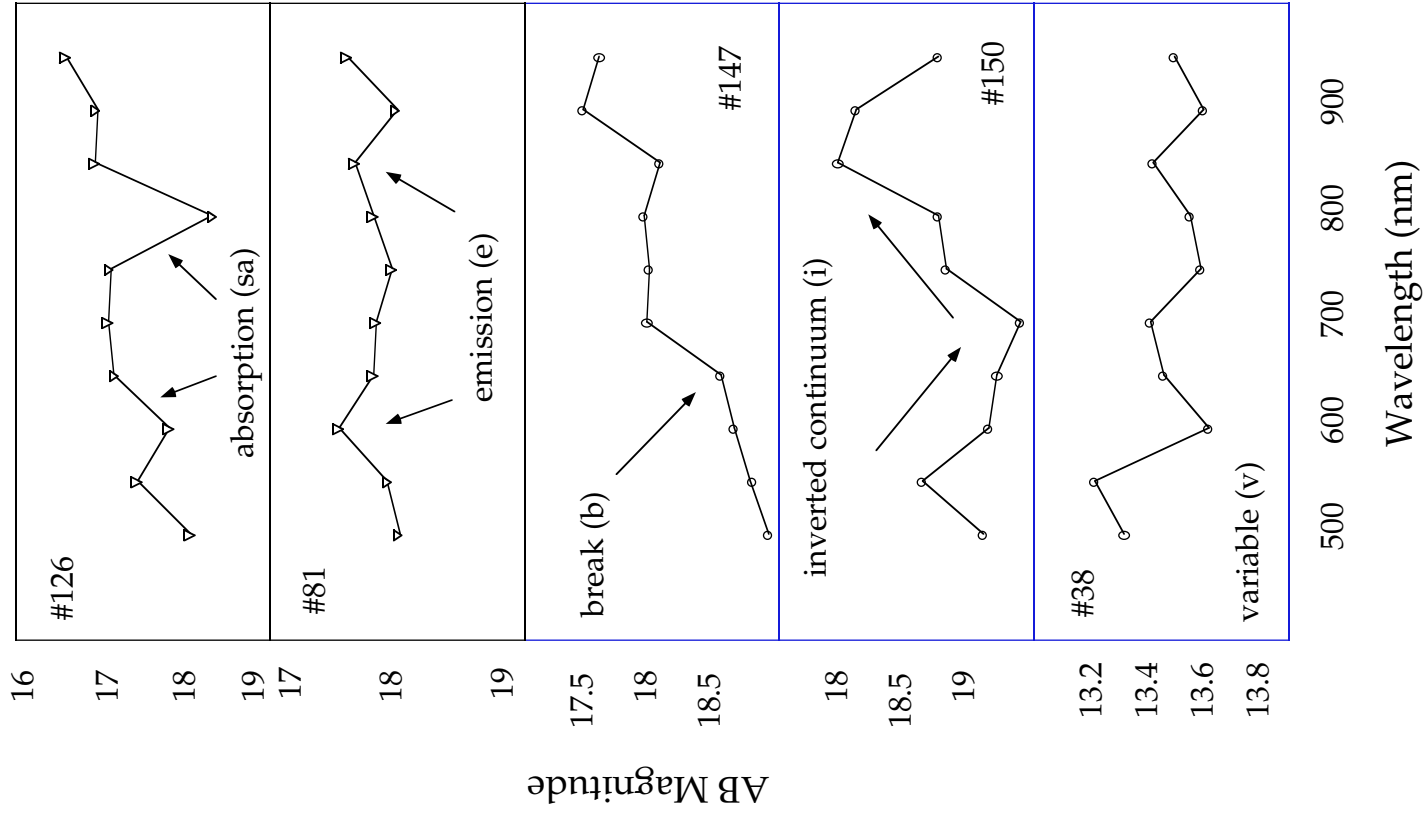

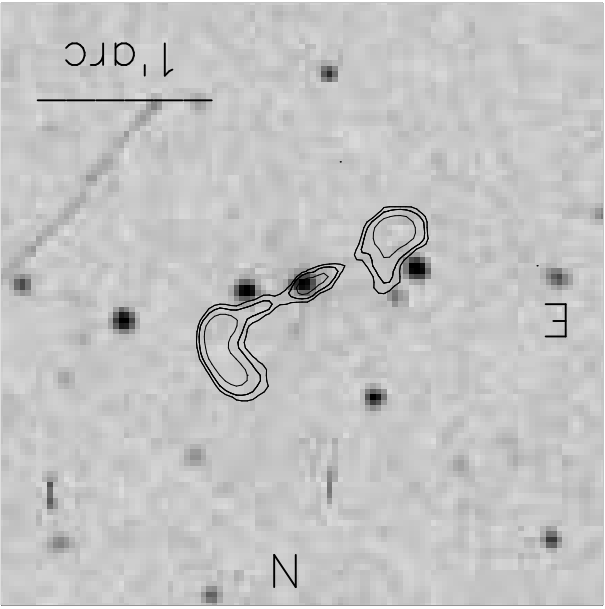 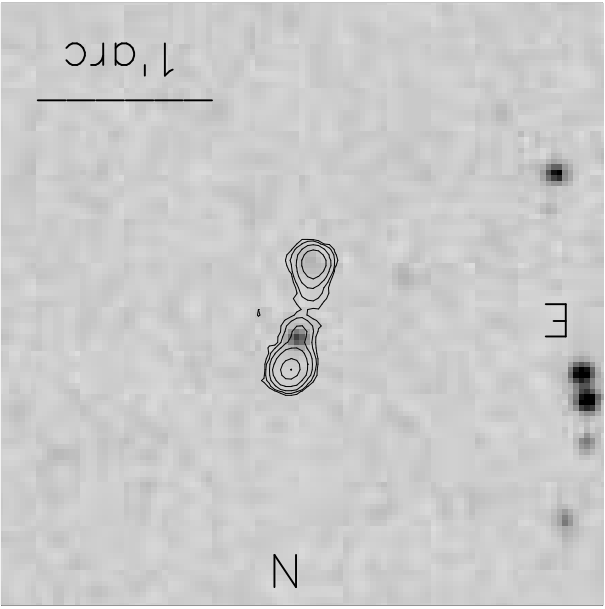

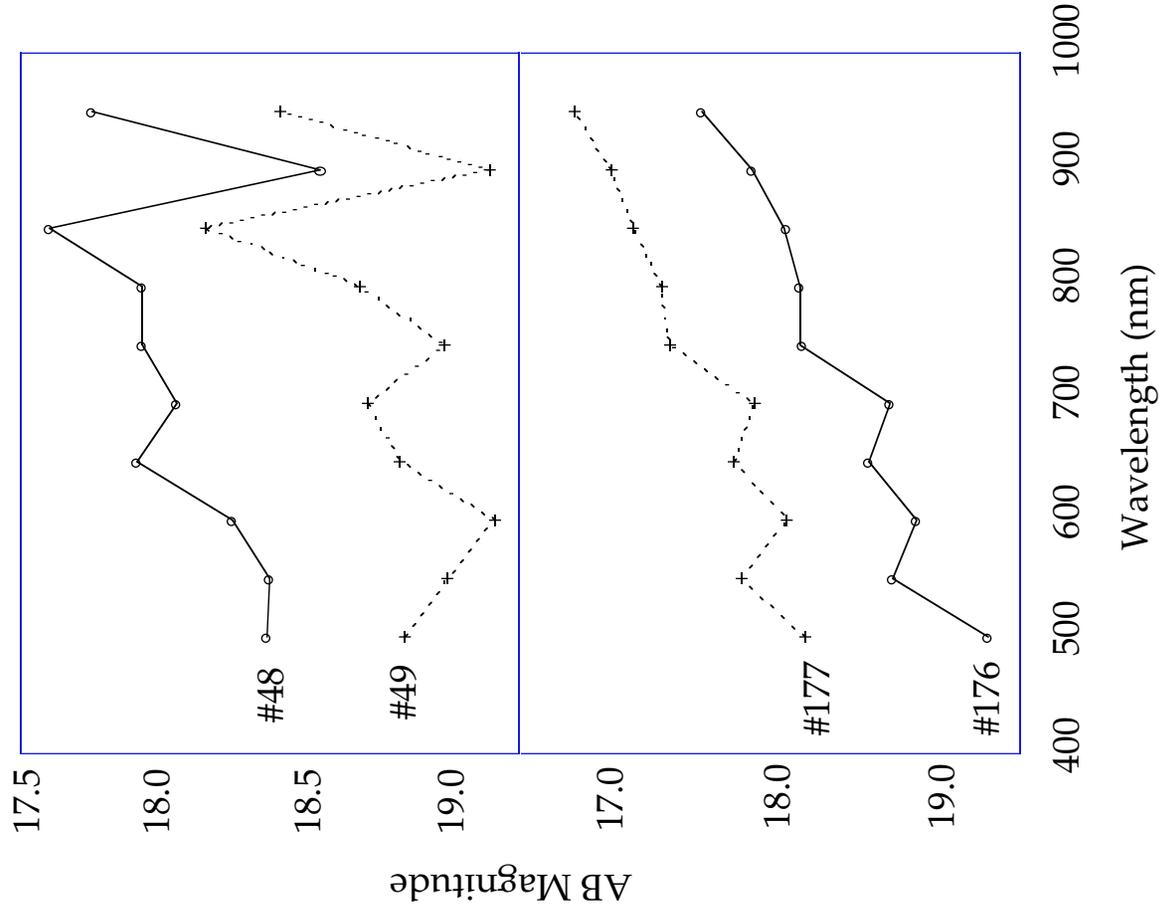

| Parameters | values |
| --- | --- |
| Solar galactocentric distance | 8 kpc |
| Disk scale length | 3.5 kpc |
| Disk Scale height | 325 pc |
| Spheroid density law | de Vaucouleurs |
| Spheroid minor/major axis | 0.8 |
| Spheroid effective radius | 2.7 kpc |
| luminosity function | globular cluster |
| No thick disk | |

| Sub-sample# | Begin - End R.A. (J2000 decimal) | Area (deg$^2$) | Central Galactic Coord. long. (°) , lat. (°) |
|---|---|---|---|
| 0-1,999 | 12.492 - 13.817 | 4.17 | 86, 83 |
| 2,000-3,999 | 13.817 - 14.882 | 3.35 | 54, 69 |
| 4,000-5,999 | 14.882 - 15.699 | 2.57 | 52, 58 |
| 6,000-7,999 | 15.699 - 16.287 | 1.85 | 53, 49 |
| 8,000-9,999 | 16.287 - 16.792 | 1.59 | 54, 42 |
| 10,000-11,999 | 16.792 - 17.169 | 1.19 | 55, 37 |
| 12,000-13,999 | 17.169 - 17.492 | 1.02 | 56, 32 |
| 14,000-15,999 | 17.492 - 17.821 | 1.26 | 58, 28 |
| 16,000-17,999 | 17.821 - 18.384 | 1.77 | 60, 23 |

| # | ra (J2000) (decimal h) | dec(J2000) (°) | m55 (AB mag) | comments•• | # | ra (J2000) (decimal h) | dec(J2000) (°) | m55 (AB mag) | comments•• |
|---|---|---|---|---|---|---|---|---|---|
| 1 | 16.723921 | 33.020596 | 16.65 | O, e65-80 | 60 | 17.113974 | 32.969524 | 18.64 | F, a65 b80 i |
| 2 | 16.116518 | 32.893772 | 15.96 | B, a60 | 61 | 15.923176 | 32.965950 | 18.25 | F, a65-90 e85 |
| 3 | 16.436195 | 32.862064 | 17.19 | B, a60 e75 | 62 | 18.290724 | 32.839352 | 16.51 | F, sa70 |
| 4 | 13.039210 | 32.804462 | 16.01 | B, a80 | 63 | 16.846828 | 33.014271 | 18.45 | F, a70 se85 |
| 5 | 17.049038 | 32.999752 | 17.19 | B, e55 a60 | 64 | 17.101196 | 32.928673 | 18.08 | F, a70 se85 |
| 6 | 17.688948 | 33.013638 | 13.76 | B, e55-80 | 65 | 17.846749 | 32.990181 | 17.66 | F, a70 v |
| 7 | 17.588503 | 32.928482 | 17.79 | B, e55-se85 | 66 | 17.939775 | 33.065960 | 18.40 | F, v |
| 8 | 15.103037 | 32.961506 | 16.83 | B, v | 67 | 16.192574 | 32.943142 | 18.43 | F, a85 |
| 9 | 16.917452 | 32.961506 | 11.11 | B, i | 68 | 13.717310 | 32.980541 | 13.68 | F, sa85 |
| 10 | 15.401999 | 32.963390 | 16.52 | B, v | 69 | 15.766602 | 32.949207 | 14.27 | F, sa90 |
| 11 | 15.777185 | 32.845936 | 13.37 | B, v | 70 | 12.495477 | 33.021168 | 14.75 | F, sa90 |
| 12 | 17.676081 | 32.874905 | 11.30 | B, i | 71 | 16.146019 | 32.965862 | 18.85 | F, b50 v |
| 13 | 13.672906 | 32.970547 | 16.11 | B, v | 72 | 15.465754 | 32.974670 | 17.72 | F, b50-60-80 |
| 14 | 14.735177 | 33.013206 | 13.81 | B, i | 73 | 17.140526 | 32.919743 | 19.05 | F, b60 sa90 |
| 15 | 12.655259 | 33.006935 | 17.24 | A, a55 | 74 | 15.725285 | 32.890343 | 18.20 | F, b65 |
| 16 | 15.888780 | 32.945999 | 18.35 | A, a60 e 85 | 75 | 14.248792 | 32.863937 | 18.32 | F, b75 a60 |
| 17 | 14.893238 | 32.913315 | 16.41 | A, a60 e85 | 76 | 14.188699 | 32.887119 | 18.27 | F, b75 a60 sa90 |
| 18 | 15.503955 | 32.875195 | 17.35 | A, a60 e85 | 77 | 15.676805 | 32.942535 | 18.39 | F, e55 |
| 19 | 12.653641 | 33.063164 | 18.22 | A, a60-85 | 78 | 16.231491 | 32.900524 | 17.86 | F, e55-80 |
| 20 | 16.809544 | 33.066399 | 16.15 | A, a65 | 79 | 17.240276 | 32.890526 | 18.12 | F, e55-85 |
| 21 | 13.773729 | 33.000942 | 17.60 | A, a65-90 v | 80 | 17.088709 | 32.972885 | 17.93 | F, se55-85 v |
| 22 | 16.131014 | 33.067314 | 17.89 | A, a90 | 81 | 15.346329 | 32.934361 | 17.93 | F, se60-85 |
| 23 | 13.796606 | 33.012356 | 18.05 | A, sa90 v | 82 | 17.156164 | 32.951824 | 18.65 | F, e65-80 |
| 24 | 16.228706 | 32.937160 | 18.10 | A, b50 | 83 | 15.860457 | 32.870056 | 16.79 | F, e65-80 |
| 25 | 13.298291 | 32.968063 | 18.37 | A, se55 a80 | 84 | 17.825848 | 32.948986 | 18.68 | F, v |
| 26 | 17.589916 | 32.949852 | 17.93 | A, e55-80 | 85 | 17.325361 | 33.036507 | 18.45 | F, se80 v |
| 27 | 17.455425 | 33.012939 | 18.93 | A, i | 86 | 16.810549 | 32.889969 | 18.07 | F, se85 |
| 28 | 15.366980 | 32.809814 | 18.65 | A, se60 | 87 | 17.073229 | 32.877617 | 18.25 | F, se85 v |
| 29 | 17.107212 | 32.913636 | 18.79 | A, e60 | 88 | 16.139938 | 33.024002 | 18.19 | F, v |
| 30 | 15.024785 | 32.885548 | 15.88 | A, a60 e65 | 89 | 17.847876 | 33.069126 | 18.90 | F, v |
| 31 | 17.578421 | 32.996418 | 15.61 | A, e65 | 90 | 17.844162 | 32.819073 | 18.06 | F, v |
| 32 | 17.297001 | 32.953693 | 18.09 | A, v | 91 | 18.230516 | 32.864479 | 18.04 | F, v |
| 33 | 15.122347 | 33.004753 | 18.57 | A, i | 92 | 18.254644 | 32.908382 | 18.20 | F, v |
| 34 | 17.514475 | 32.954876 | 18.50 | A, v | 93 | 16.146976 | 32.980961 | 19.00 | F, v |
| 35 | 15.913959 | 32.903946 | 16.72 | A, v | 94 | 18.000252 | 32.946537 | 14.29 | F, v |
| 36 | 14.884324 | 32.860538 | 18.55 | A, v | 95 | 18.202223 | 33.004513 | 18.73 | F, v |
| 37 | 16.211849 | 33.017284 | 17.99 | A, v | 96 | 17.670643 | 33.027397 | 15.75 | F, v |
| 38 | 17.546465 | 32.934525 | 13.19 | A, v | 97 | 17.356752 | 32.892448 | 18.08 | F, v |
| 39 | 16.922382 | 32.917831 | 17.83 | A, v | 98 | 17.427036 | 32.976620 | 18.26 | F, v |
| 40 | 15.432915 | 32.922531 | 18.43 | A, v | 99 | 17.055372 | 32.935490 | 18.79 | F, e85 v |
| 41 | 17.121136 | 32.971241 | 17.55 | A, v | 100 | 17.329021 | 33.024418 | 18.41 | F, v |
| 42 | 15.207924 | 33.065437 | 16.90 | A, v | 101 | 16.708237 | 32.950710 | 18.09 | F, v |
| 43 | 15.386465 | 33.037392 | 16.63 | A, v | 102 | 17.627253 | 33.025131 | 18.60 | F, v |
| 44 | 16.025953 | 32.978031 | 18.74 | F, a60 e80 | 103 | 16.214687 | 32.997471 | 18.77 | F, v |
| 45 | 13.416241 | 32.957035 | 17.71 | F, a60 e85 | 104 | 16.598019 | 33.016251 | 18.61 | F, e85 v |
| 46 | 16.211849 | 32.934746 | 18.11 | F, a60 e85 | 105 | 16.272659 | 33.025818 | 17.91 | F, b50 v |
| 47 | 17.546595 | 33.009304 | 17.08 | F, a60-90 | 106 | 15.120415 | 32.892814 | 15.75 | F, v |
| 48 | 16.123663 | 32.870831 | 18.40 | F glq, e65-85 a90 | 107 | 15.167593 | 32.892448 | 18.08 | F, v |
| 49 | 16.123491 | 32.876652 | 18.97 | F glq, v | 108 | 15.925856 | 32.939995 | 17.63 | F, e85 v |
| 50 | 12.679251 | 33.063791 | 17.93 | F ocrs, a85 v | 109 | 15.117019 | 32.994129 | 18.74 | F, v |
| 51 | 13.939829 | 32.928066 | 18.25 | F, sa60 | 110 | 15.361760 | 33.014713 | 18.18 | F, v |
| 52 | 14.165607 | 33.034233 | 17.69 | F, a60 e80 | 111 | 15.927396 | 32.978886 | 18.98 | F, v |
| 53 | 14.817472 | 32.911240 | 18.13 | F, a60 e85 | 112 | 15.945646 | 32.933601 | 18.41 | F, v |
| 54 | 14.814044 | 32.878841 | 17.98 | F, a60 e85 | 113 | 13.747837 | 32.948402 | 18.56 | F, v |
| 55 | 14.980004 | 32.912010 | 19.01 | F, a60-90 | 114 | 16.125002 | 33.053285 | 18.87 | F, v |
| 56 | 14.366044 | 33.056713 | 18.45 | F, a60-90 | 115 | 15.888523 | 32.905727 | 17.64 | F, v |
| 57 | 15.924702 | 32.890377 | 18.32 | F, a60-90 b75 | 116 | 15.847326 | 33.024418 | 17.96 | F, sa60 |
| 58 | 15.912476 | 33.062706 | 18.37 | F, a65 | 117 | 15.690289 | 32.937298 | 18.58 | F, v |
| 59 | 15.820074 | 32.894077 | 18.21 | | 118 | 15.085055 | 32.999649 | 17.17 | F, v |

| # | ra (J2000) (decimal h) | dec(J2000) (°) | m55 (AB mag) | comments•• | # | ra (J2000) (decimal h) | dec(J2000) (°) | m55 (AB mag) | comments•• |
|---|---|---|---|---|---|---|---|---|---|
| 119 | 15.437516 | 32.869705 | 18.49 | F, v | 178 | 14.869380 | 32.969631 | 17.80 | K, a60-70 |
| 120 | 16.003151 | 32.986099 | 18.77 | F, v | 179 | 16.842138 | 33.021889 | 17.69 | K, a70 |
| 121 | 15.562891 | 32.935043 | 18.65 | F, i | 180 | 16.621000 | 32.991989 | 18.38 | K, a70 |
| 122 | 16.639601 | 32.853397 | 17.81 | G ocrs, v | 181 | 15.967205 | 32.891788 | 17.69 | K, a70 |
| 123 | 13.755824 | 32.939354 | 18.56 | G, a60 a75 | 182 | 16.404322 | 32.889938 | 18.80 | K, a70 e85 |
| 124 | 15.345398 | 32.989735 | 18.49 | G, sa60 a75 | 183 | 14.867328 | 32.938610 | 17.96 | K, sa70 v |
| 125 | 15.018947 | 32.879272 | 18.47 | G, a60 se85 | 184 | 13.709273 | 33.036926 | 19.13 | K, sb55 |
| 126 | 15.392588 | 32.848072 | 17.44 | G, a60 sa80 | 185 | 16.654186 | 32.907402 | 16.35 | A, a60 v |
| 127 | 14.692578 | 32.927032 | 18.36 | G, a60 sa85 | 186 | 16.288073 | 32.974670 | 18.39 | K, b50 e85 |
| 128 | 17.352865 | 32.939838 | 18.10 | G, v | 187 | 12.519277 | 32.948395 | 18.61 | K, sb60 |
| 129 | 17.342436 | 33.059921 | 18.06 | G, v | 188 | 12.887023 | 33.034122 | 17.77 | K, b70 |
| 130 | 18.304464 | 33.014500 | 18.44 | G, v | 189 | 16.900234 | 32.883560 | 18.44 | K, b70 e55 |
| 131 | 17.849007 | 32.921776 | 18.52 | G, a70 se85 | 190 | 16.397871 | 33.053730 | 18.35 | K, b70 e85 |
| 132 | 17.870338 | 33.023129 | 19.29 | G, a70-85 | 191 | 17.490679 | 32.896881 | 18.67 | K, se85 |
| 133 | 17.336882 | 32.893444 | 18.77 | G, v | 192 | 17.443306 | 32.929081 | 18.06 | K, e55 se85 |
| 134 | 17.340992 | 33.035114 | 18.26 | G, a70 i | 193 | 16.876751 | 33.043118 | 18.30 | K, se55 b70 |
| 135 | 15.823177 | 32.931190 | 18.43 | G, b50 a65 e80 | 194 | 16.952143 | 32.934662 | 19.41 | K, e65-85 |
| 136 | 15.934926 | 32.940643 | 17.83 | G, b50-75 | 195 | 16.983458 | 33.016548 | 19.11 | K, e65-85 |
| 137 | 13.887213 | 32.944538 | 19.14 | G, b55 | 196 | 16.944590 | 32.922012 | 18.84 | K, e65-85 |
| 138 | 14.002308 | 33.069347 | 19.02 | G, sb55 b80 | 197 | 16.913128 | 32.977657 | 18.71 | K, e65-85 |
| 139 | 13.235456 | 33.040138 | 16.39 | G, b55-80 | 198 | 16.721235 | 33.012829 | 15.15 | K, e65-85 |
| 140 | 13.908546 | 32.950611 | 19.05 | G, sb55 b80 | 199 | 12.583992 | 32.959026 | 19.10 | K, e70-90 |
| 141 | 13.817575 | 33.065121 | 17.50 | G, b60 | 200 | 16.255863 | 32.931995 | 18.24 | K, se85 |
| 142 | 13.864271 | 33.056900 | 18.70 | G, b60 | 201 | 17.377533 | 32.992626 | 18.21 | K, v |
| 143 | 17.169315 | 33.069931 | 16.23 | G, b60 e65 | 202 | 16.267193 | 32.917110 | 18.38 | K, v |
| 144 | 16.126150 | 32.935555 | 17.62 | G, b60 a90 | 203 | 13.543396 | 32.896542 | 18.14 | M gc, b70 |
| 145 | 17.163759 | 32.889637 | 18.80 | G, b60 se85 | 204 | 17.387367 | 33.034721 | 18.67 | M, b55 a70-85 |
| 146 | 15.916878 | 32.884407 | 18.67 | G, b60 sa90 | 205 | 17.097330 | 33.012352 | 18.83 | M, b70 e60 |
| 147 | 12.601104 | 32.992477 | 12.73 | G, sb65-85 | 206 | 12.573947 | 32.889343 | 18.52 | M, b55-75 |
| 148 | 17.317432 | 32.943245 | 18.78 | G, b70 e55-65 | | | | | |
| 149 | 16.042187 | 33.000416 | 18.42 | G, b70 e65 | | | | | |
| 150 | 17.086172 | 33.059448 | 17.67 | G, i | | | | | |
| 151 | 12.812544 | 32.885414 | 18.62 | G, b85 | | | | | |
| 152 | 16.134058 | 32.933578 | 18.91 | G, e55-85 | | | | | |
| 153 | 16.994238 | 32.873211 | 18.25 | G, e55 se85 | | | | | |
| 154 | 16.618204 | 32.894547 | 18.34 | G, e65 | | | | | |
| 155 | 16.025713 | 32.821509 | 15.53 | G, e65 se85 | | | | | |
| 156 | 15.972535 | 32.884407 | 18.67 | G, e75 b95 | | | | | |
| 157 | 17.004301 | 32.992477 | 18.15 | G, sb65-85 | | | | | |
| 158 | 17.170305 | 32.936333 | 18.53 | G, se85 v | | | | | |
| 159 | 17.150970 | 32.971741 | 18.53 | G, e85 v | | | | | |
| 160 | 17.478706 | 32.891472 | 18.71 | G, e85 v | | | | | |
| 161 | 12.500440 | 33.003380 | 19.01 | G, e90 | | | | | |
| 162 | 17.482128 | 33.044647 | 19.46 | G, sb55 e90 | | | | | |
| 163 | 14.427979 | 32.864418 | 18.58 | G, v | | | | | |
| 164 | 17.380459 | 32.894547 | 17.94 | G, v | | | | | |
| 165 | 12.917326 | 32.947987 | 18.58 | G, v | | | | | |
| 166 | 13.757835 | 32.901485 | 19.07 | G, v | | | | | |
| 167 | 17.436001 | 32.866089 | 17.94 | G, b60 v | | | | | |
| 168 | 18.308521 | 32.887566 | 18.21 | G, v | | | | | |
| 169 | 16.245291 | 33.025639 | 18.20 | G, v | | | | | |
| 170 | 16.243963 | 33.042168 | 18.70 | G, v | | | | | |
| 171 | 15.621979 | 33.032482 | 16.34 | G, v | | | | | |
| 172 | 16.668447 | 32.813549 | 18.23 | G, v | | | | | |
| 173 | 15.670748 | 33.064629 | 18.34 | G, v | | | | | |
| 174 | 17.163877 | 32.895538 | 18.35 | G, v | | | | | |
| 175 | 15.053413 | 32.929394 | 18.89 | G, v | | | | | |
| 176 | 17.386164 | 33.015015 | 18.54 | G, v | | | | | |
| 177 | 17.386444 | 33.027573 | 18.74 | K glq, v | | | | | |
| | 17.386444 | 33.030758 | 17.82 | K glq, v | | | | | |

•• OBAFGKM=Gunn&Stryker best-fitted spectrum, s=strong, a=absorption, e=emission, b=break, v=variable, glq=gravitational lense candidate, gc=galaxy cluster, ocrs=optical counterpart of radio source.